\documentclass[prd,a4paper,twocolumn,nofootinbib]{revtex4}
\usepackage{graphicx}
\begin{document}

\title{Neutrino Mass, Dark Energy, and the Linear Growth Factor}
\author{Angeliki Kiakotou$^1$, {\O}ystein Elgar{\o}y$^2$, Ofer Lahav$^1$}
\affiliation{$^1$ Department of Physics and Astronomy, University College London, 
Gower Street, London, WC1E 6BT, UK\\
$^2$ Institute of Theoretical Astrophysics, University of Oslo,Box 1029, 0315 Oslo, NORWAY\\}

\date{\today}

\begin{abstract}

We study the degeneracies between neutrino mass and dark energy as they manifest themselves in cosmological observations. In contradiction to a popular formula in the literature, the suppression of the matter power spectrum caused by  massive neutrinos is not just a function of the ratio of neutrino to total mass densities $f_{\nu}=\Omega_{\nu}/\Omega_{m}$, but also each of the densities independently. We also present a fitting formula for the logarithmic growth factor of perturbations in a flat universe, $f(z, k;f_\nu,w,\Omega_{\rm DE}) \approx [1-A(k)\Omega_{DE}f_{\nu}+B(k)f_{\nu}^{2}-C(k)f_{\nu}^{3}] \Omega_{m}^{\alpha}(z)$, where $\alpha$ depends on the dark energy equation of state parameter $w$. We then discuss  cosmological probes where the $f$ factor directly appears: 
peculiar velocities, redshift distortion and the Intergrated Sachs-Wolfe effect. 
We also modify the approximation of Eisenstein \& Hu (1999) for the power spectrum of fluctuations in the presence of massive neutrinos and provide a revised code\footnote {http://www.star.ucl.ac.uk/$\sim$lahav/nu\_matter\_power.f}.

\end{abstract}
\maketitle

\section{Introduction}   

The latest results from the WMAP satellite \cite{wmap3param} confirm 
the success of the $\Lambda$CDM model, where $\sim$ 75 \% of the 
mass-energy density is in the form of dark energy, and matter, most 
of it in the form of Cold Dark Matter (CDM) making up the remaining 
25 \%.  Neutrinos with masses on the eV scale or below will be a hot 
component of the dark matter and will free-stream out of overdensities and 
thus wipe out small-scale structures.  This fact makes it possible to 
use observations of the clustering of matter in the universe to put 
upper bounds on the neutrino masses.  A thorough review of the 
subject is found in \cite{pastorreview}.  With the improved quality 
of cosmological data seen in recent years, the upper limits have improved, 
and some quite impressive claims have been made in the recent literature, e.g. \cite{seljak}. 
 
Present cosmological neutrino mass limits make use of the suppression 
effect of the neutrino free-streaming at a fixed, given redshift.  
As our ability to map out the mass distribution at different epochs 
of the cosmic history improves, by doing, e.g., weak lensing tomography, we will gain sensitivity by in addition using the effect of massive neutrinos on the growth rate of density fluctuations.  One key issue which then arises is possible degeneracies between neutrino masses and cosmological parameters. 
In this paper we focus on the degeneracy between the dark energy equation 
of state and the neutrino masses.  We show that the combined effect of 
neutrinos and dark energy can be parametrized in a simple manner, and 
that the degeneracy can be broken by mapping out 
the large-scale structure over a reasonably wide range of redshifts. We also explore the effect of massive neutrinos on the matter power spectrum P(k), and we modify the formula given by Eisenstein \& Hu \cite{eisenstein}, such that it would be valid for realistic properties of massive neutinos. The dependencies of three important ingredients of the universe on the equation of state parameter $w$ and the neutrino density $\Omega_{\nu}$ are shown in Table \ref{table:geometry}.

The outline of the paper is as follows.
 In Section II we contrast common approximations to the power spectrum of fluctuations 
 with the exact results from CAMB, and we provide a new modified approximation.
 In Section III we provide a new fitting formula for  the linear theory growth of perturbations 
 in the presence of massive neutrinos.
 In section IV we discuss the parameter degeneracy between dark energy parameters
 and neutrino masses, and we illustrate how it manifests itself in peculiar velocities and 
 the Integrated Sachs Wolfe effect.  Our conclusions are summarized in Section V.
\par

\begin{table}
\begin{tabular}{|c|c|c|}
\hline

 & $\nu$ mass & dark energy \\
\hline
Geometry & $\times$ & $\surd$\\
Matter Power Spectrum $P(k,z=0)$ & $\surd$ & $\times$\\
Linear Growth Function $\delta (z)$ & $\surd$ & $\surd$\\
\hline
\end{tabular}
\label{table:geometry}
\caption{The ability of observational tests based on geometry, matter power spectrum and linear growth to probe neutrino mass and dark energy. In the text we discuss the sensitivity of $P(k,z)$ to neutrino mass.}
\end{table}

\section{Massive neutrinos and the matter power spectrum}

\begin{figure}[htbp]
\includegraphics[width=9cm, height=6cm]{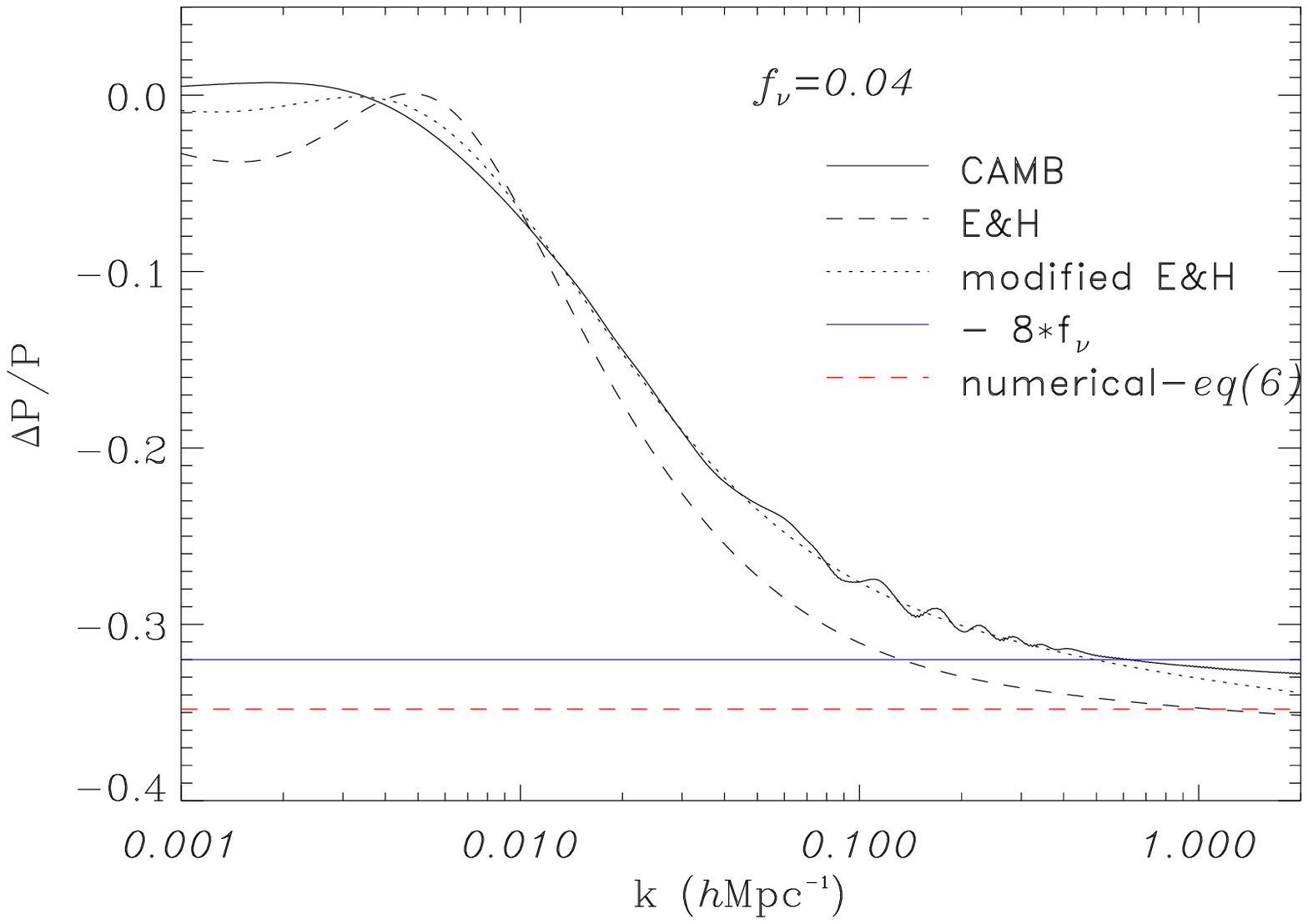}
\includegraphics[width=9cm, height=6cm]{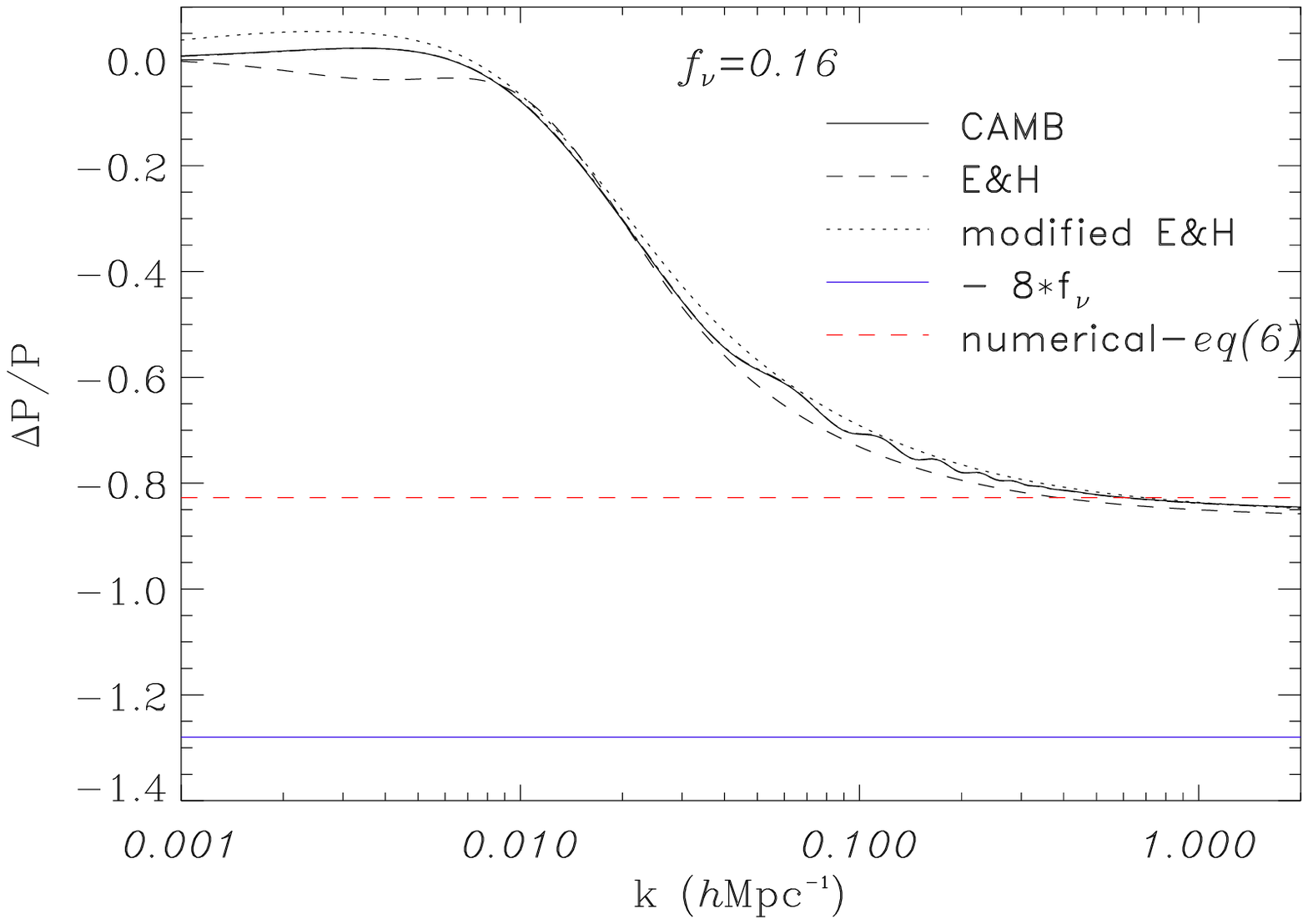}
\caption{The quantity $\Delta P/ P$ at $z=0$ defined in equation (\ref{eq:deltap1}), derived using CAMB for fixed $w= -1, \Omega_{m}=0.25, \Omega_{b}=0.04, h=0.7$ (solid black line). The upper plot is for $f_{\nu}=0.04$ and the lower plot for $f_{\nu}=0.16$. For comparison we show the fitting formula from Eisenstein \& Hu \cite{hu0} (dashed black line), and our modification to their formula (dotted black line). The horizontal blue line is the approximation $- 8 f_{\nu}$ from \cite{hu0}, and the horizontal dashed red line is our numerical solution to equation (\ref{eq:eq2}).}
\label{fig:power}
\end{figure}

Most cosmological neutrino mass limits make use of the matter power 
spectrum $P(k)$ in some guise, although it is also possible to obtain a 
limit from cosmic microwave background data alone, see \cite{ichikawa,ichikawa2,kristiansen}.\\
%Since cosmological perturbations cannot grow significantly in a radiation dominated universe, an important parameter is the time of equality between the densities of matter and radiation
%\begin{equation}
%z_{eq}= 23\,900\;\Omega_m h^2 (T/2.726\textrm{K})^{-4}\;-\;1.
%\label{eq:zeq}
%\end{equation}
%The scale of the particle horizon at this epoch, 
%\begin{equation}
%k_{eq}=4.7 \,\, \textrm{x} \; 10^{-4} \sqrt{1+z_{eq}}\; \textrm{h/Mpc}
%\label{eq:keq}
%\end{equation}
%is imprinted in the matter transfer function: perturbations on smaller scales $(k > k_{eq})$ can only start growing after $z_{eq}$, while those on larger scales $(k < k_{eq})$ keep growing at any time. This leads to the suppression of the transfer function at $k > k_{eq}$, and hence of the matter power spectrum.
\\
A useful way to consider the effect of neutrino mass on the power spectrum is to 
consider the following quantity at a given redshift:
 \begin{equation}
\frac{\Delta P(k)}{P(k)} = \frac{P(k;f_\nu)-P(k;f_\nu=0)}{P(k;f_\nu=0)}.
\label{eq:deltap1}
\end{equation}
A common heuristic explanation for the role of the matter power spectrum 
in deriving neutrino mass limits is the approximate expression   
$\frac{\Delta P(k)}{P(k)} \approx -8f_\nu$,
describing the suppression of small-scale power caused by neutrino free-streaming, 
a result valid for $f_\nu \ll 1$, and first given in \cite{hu0}. 
See \cite{max1,pastorreview} for derivations. In Figure (\ref{fig:power}), we compare this approximation to the matter power spectrum using CAMB \footnote{Accuracy of CAMB is set to 0.3 \%.} \cite{camb}, but also to the Eisenstein \& Hu approximation \cite{eisenstein}, to our modification to this E\&H approximation and to the numerical solution given by equation (\ref{eq:eq2}), for $f_\nu=\frac{\Omega_\nu}{\Omega_m}=0.04$ and $f_\nu=0.16$. An approximation for $P(k,z)$ is given in \cite{pastorreview}
 \begin{equation}
P(k,z)=\left\{
{\rm \begin{tabular}{llr}
$\displaystyle \left(\frac{g(z)}{(1+z)g(0)}\right)^2 \!\!P(k,0)$
&~for&
$aH < k < k_{\rm nr}$~, \\
$\displaystyle \left(\frac{g(z)}{(1+z)g(0)}\right)^{2p} 
\!\;P(k,0)$ 
&~for&
$k \gg k_{\rm nr}$ ~,
\end{tabular}}
\right.
\label{eq:julien}
\end{equation}
where $p\approx 1- 3/5 \, f_{\nu}$ based on equation (\ref{eq:deltap}) below. We give a better approximation to this factor in Section \ref{sec:lambda}. The non-relativistic scale, $k_{nr}$, is given by
\begin{equation}
k_{nr}\approx 0.018 \Omega_m ^{1/2}\;(\frac{m}{1\textrm{eV}})^{1/2}\; h\; \textrm{Mpc}^{-1}.
\label{eq:knr}
\end{equation}
For another approximation valid for Mixed Dark Matter see \cite{durr}.

As a rule of thumb, the present-epoch matter power spectrum is considered 
to be in the linear regime for comoving wavenumbers $k<0.10$--$0.15\;h\,{\rm Mpc}^{-1}$, 
and we see from Figure (\ref{fig:power})  that $\Delta P/P$ obtained from CAMB tends to a 
constant only for $k>1.5\ \textrm{Mpc}^{-1}$, which is well into the 
non-linear regime of structure formation. This fact is also evident 
from Figures (12) and (13) in \cite{pastorreview}.
Thus, $\frac{\Delta P}{P} \approx -8 f_\nu$ should only be used as a heuristic guide 
to the effect of massive neutrinos on the power spectrum since it's not valid at the length scales where the matter power spectrum can be said to be in the linear regime.   
Furthermore,  we note that, as expected, it works well only for very small $f_{\nu}$ whereas for large neutrino masses this approximation breaks down.
Moreover, Figure (\ref{fig:power}) also shows that for small neutrino masses the E\&H approximation breaks down. However, by modifying the master transfer function used by E\&H, we managed to minimise the error between CAMB and the E\&H approximation for $f_{\nu}=0.04$, from $20\% $ to only $3\%$ for $0.02<k<0.15\; h\ \textrm{Mpc}^{-1}$.  This modification works much better for small neutrino masses than its predecessor used to, as well as for very massive neutrinos. For $0.15 \leq \Omega_m \leq 0.8, \Omega_b / \Omega_m
\leq 0.3, f_\nu \leq 0.3, z=0$ and $N_\nu =3$ the accuracy of the fitting formula is quite high. Note that the formula works equally well for $\Omega_\nu =0.$
Our revised code can be downloaded from http://zuserver2.star.ucl.ac.uk/$\sim$lahav/nu\_matter\_power.f.
\par

We also explore the effect of the dark energy equation of state ($w$) on the matter power spectrum. For a constant $f_{\nu}$ and in the regime of interest $0.02<k<0.15\; h\ \textrm{Mpc}^{-1}$, $w$ affects the growth factor $\delta$ which affects the matter power spectrum by changing the amplitude, but this can be altered by taking the ratio $\frac{\Delta P}{P}$ where this effect is minimal, i.e. is independent of $w$ (to 4\% in the regime of interest).
 \par

For a fixed value of the parameter combination $\Omega_{\rm m}h$, the 
impact of massive neutrinos on the matter power spectrum is controlled 
by  the fractional contribution of neutrinos to the total
 mass density in the Universe, ie. 
$f_{\nu}=\frac{\Omega_{\nu}}{\Omega_{m}}$ 
The scale where the suppression of power from 
neutrino free-streaming sets in is controlled by the comoving Hubble radius 
at the time when the neutrinos became non-relativistic, corresponding 
to a comoving wavenumber 
\begin{equation}
k_{fs}=0.10\Omega_{m} \; h \sqrt{f_{\nu}}.
\label{eq:kfs}
\end{equation}

\begin{figure}[htbp]
\includegraphics[width=9cm,height=6cm]{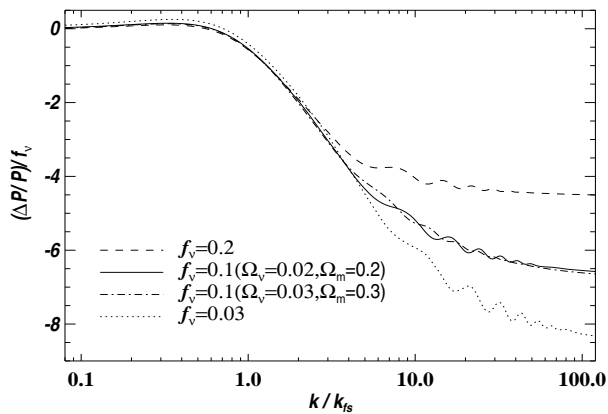}
\caption{The dependence of $\Delta P/P$ at $z=0$ on $\Omega_{m}$ and $\Omega_{\nu}$, illustrating, via the scaling by $k_{fs}$, equation (\ref{eq:kfs}),  that just the ratio $f_{\nu}=\frac{\Omega_{\nu}}{\Omega_{m}}$ is insufficient to fully parametrize $\Delta P /P$.
From top to bottom: $f_{\nu}$=0.2 (dashed line), 2 models with $f_{\nu}$=0.1($\Omega_{\nu}$=0.03, $\Omega_{m}$=0.3 (dash-dotted line), $\Omega_{\nu}$=0.02, $\Omega_{m}$=0.2 (solid line)), $f_{\nu}$=0.03 (dotted line). We see that the ratio $\frac{\Delta P}{P}/ f_{\nu}$ tends to a constant only for $k>50k_{fs}$, but not to a universal constant.}
\label{fig:kfs}
\end{figure}

Note the obvious degeneracy with $\Omega_{\rm m}h$.  
This parameter, in models with negligible baryon density, sets the 
scale of the Hubble radius at matter-radiation equality, and so it 
determines the scale at which the matter power spectrum bends over in 
the case of massless neutrinos.  For realistic baryon densities, there 
is an additional dependence on $\Omega_{\rm b}$. 
There is a statement, sometimes found in the literature, 
that the power spectrum depends only on $f_\nu$
and not the indepenent values of the neutrino and matter densities. 
This is only true in the case where $\Omega_{\rm m}h$ and $\Omega_{\rm b}$ 
are held fixed as $f_\nu$ varies, and only for $k>0.8\; h\ \textrm{Mpc}^{-1}$, 
since the neutrino free-streaming also enters this picture. However, 
this can be altered by normalising $k$ by $k_{fs}$ (free-streaming scale). 
As shown in Figure (\ref{fig:kfs}), $\frac{\Delta P}{P}$ is a function of the ratio $f_{\nu}$ and $\frac{k}{k_{fs}}$. Moreover, we see that $(\Delta P/ \; P)/\; f_{\nu}$ tends to -8 only for small neutrino masses whereas it goes to -4.5 for $f_{\nu}=0.2$. \par

\begin{figure} 
\includegraphics[width=9cm,height=5cm]{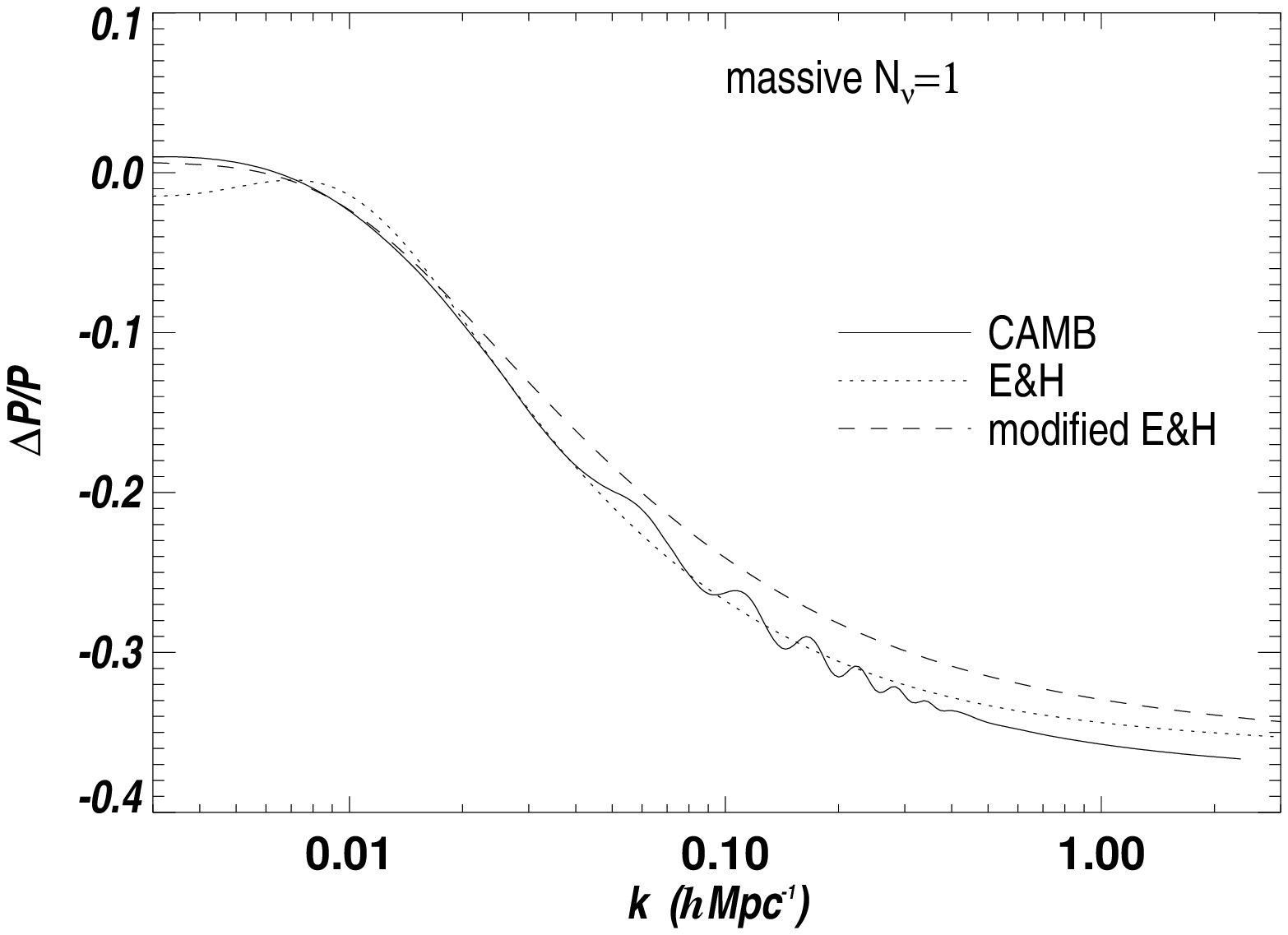}
\includegraphics[width=9cm,height=5cm]{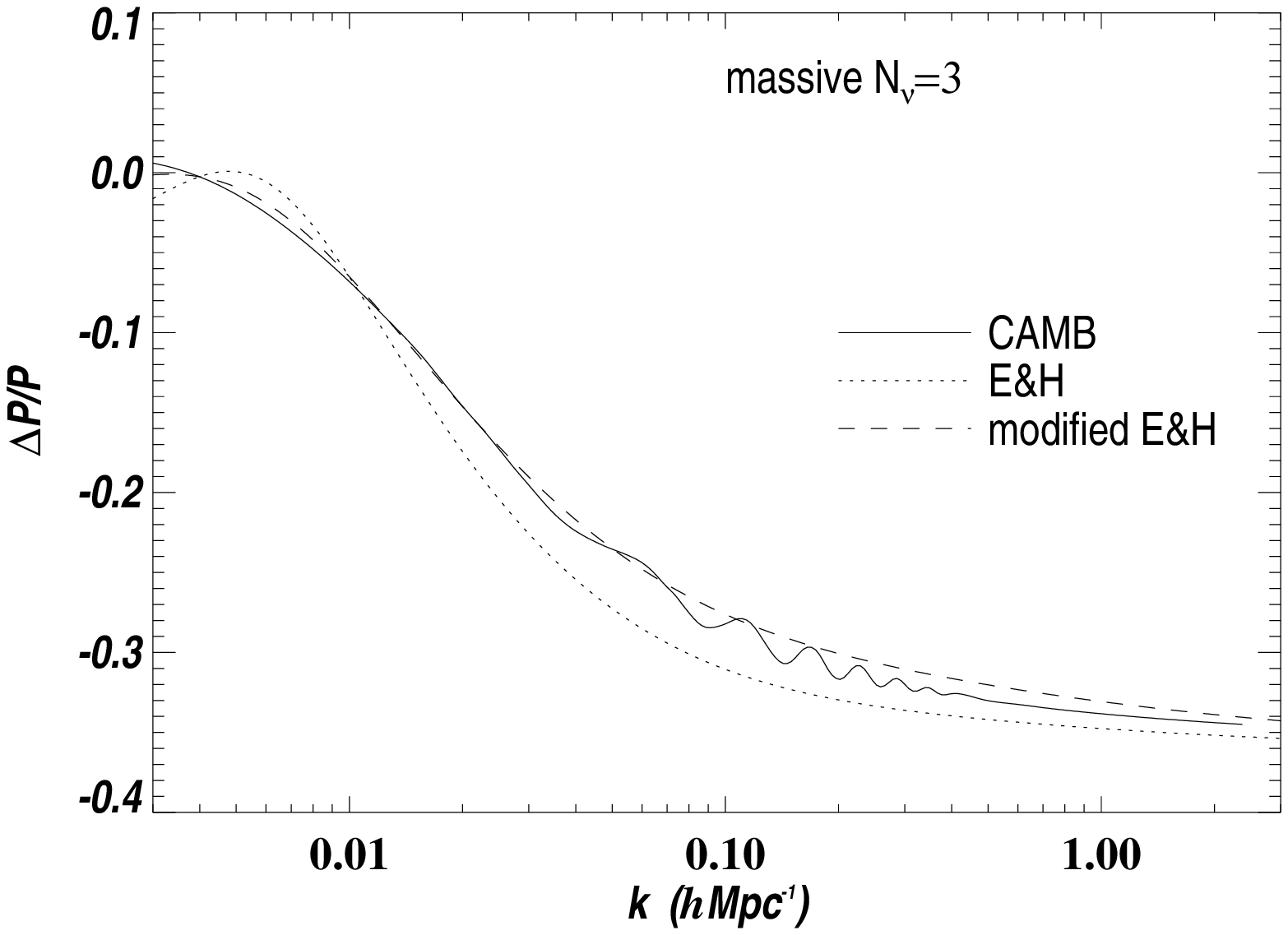}
\caption{$\Delta P / P$ at $z=0$ for 1 massive, 2 massless neutrinos (top panel) and 3 massive neutrinos (bottom panel), using CAMB (full line), E\&H (dotted line),and our modified fitting formula (dashed line).In all cases $f_{\nu}=0.04, \Omega_m =0.25, \Omega_b=0.04$.}
\label{fig:nunum}
\end{figure}

Exploring the E\&H approximation to CAMB in Figure (\ref{fig:nunum}), we observe that their formula provides best results for only 1 massive neutrino and 2 massless neutrinos with large $f_{\nu}$, whereas there is considerably less power on small scales for 3 massive neutrinos. However with our modified formula, this effect is altered and the approximation is now valid for 3 massive neutrinos of any mass.

\section{Linear growth: an analytical approximation}
\label{sec:growth}

The equation for linear evolution of density perturbations is

\begin{equation}
\ddot{\delta} + 2\frac{\dot{a}}{a}\dot{\delta} = 4\pi G \rho_0\delta,
\label{eq:eq1}
\end{equation}
where $\delta = \delta\rho_{\rm m}/\rho_{\rm m}$, and $\rho_{\rm m}$ 
and $\delta \rho_{\rm m}$ is the density and the overdensity of 
matter, respectively.  Light, massive neutrinos inhibit structure 
formation on small scales because they free-stream out of the 
dark matter potential wells.  Roughly, this can be taken into account by 
by multiplying 
the driving term on the right-hand side of equation (\ref{eq:eq1}) 
by a factor  $\frac{\Omega_{\rm cdm}}{\Omega_{\rm cdm}+\Omega_{\nu}}
=\frac{\Omega_{\rm cdm}}{\Omega_{\rm m}} = 1-f_\nu$, $f_{\nu} = \frac{\Omega_{\nu}}{ \Omega_{\rm m}}$. This equation is valid for $k>0.2  \;h{\textrm{ Mpc} ^{-1}}$.

\begin{equation}
\ddot{\delta} + 2\frac{\dot{a}}{a}\dot{\delta} = 4\pi G \rho_0(1-f_\nu)\delta.
\label{eq:eq2}
\end{equation}

\subsection{Einstein De Sitter Universe}
For an Einstein de Sitter universe,the solution to equation (\ref{eq:eq2}) is  given by \cite{pastorreview}, \cite{bond}, \cite{maxnon}

\begin{equation}
\delta \propto a^{p},
\label{eq:deltap}
\end{equation}
 where $p \approx 1-\frac{3}{5} f_{\nu}$.
 Some authors (e.g. \cite{maxnon}), have used this relation to estimate crudely the suppression of the power spectrum due to massive neutrinos \footnote{The qualitative derivation for the suppression of $P(k)$ goes as follows. From equation (\ref{eq:deltap}), the growth from matter-radiation equality epoch $a_{eq}$ to the present $a_0$ is
 \begin{equation}
 \frac{\delta (a_{0})}{\delta (a_{eq})}= (1+z_{eq})(1+z_{eq})^{-\frac{3}{5}f_{\nu}} =(1+z_{eq}) \rm{e}^{-\frac{3}{5} f_{\nu}\ln(1+z_{\rm eq})}.
 \end{equation}
The power spectrum $P (k)$ is the variance of the fluctutions $\delta$ in Fourier space, so massive neutrinos suppress it by the same factor as it suppresses $\delta ^{2}$, i.e.
 \begin{equation}
 \frac{P(k,f_{\nu})-P(k,f_{\nu}=0)}{P(k,f_{\nu}=0)}\simeq -\frac{6}{5} f_{\nu}\ln(1+z_{\rm eq}).
 \label{eq:fnu96}
 \end{equation}
Conceptually this derivation contrasts two scenarios (with and without massive neutrinos) which yield at the present epoch the same amplitude of fluctuations. It also assumes that $\Omega_m$ is the same for both scenarios, hence $(1+z_{eq}) \approx23900\Omega_{m}h^2$. For the concordance model $\Omega_m h^2 = 0.175$. This gives for the RHS of equation (\ref{eq:fnu96}), $\frac{\Delta P}{P} \approx -9.6 f_{\nu}$.
Different coefficients may be obtained by taking into account 
whether the neutrinos became non-relativistic before or after matter-radiation 
equality. } to be $\frac{\Delta P}{P} \approx -8 f_{\nu}$.
However, as we showed in the previous section this is 
a poor approximation.
Our Figure \ref{fig:power} shows the limitation of this approach as it is only valid for very large scales $k>0.5\;h\,{\rm Mpc}^{-1}$ and small neutrino masses $f_\nu<0.05$. More importantly, we have shown in Figure \ref{fig:kfs} that the suppression is not just a function of the ratio $f_\nu$, but of the matter and neutrino densities separately.

\begin{figure}[tp]
\includegraphics[width=9cm,height=5cm]{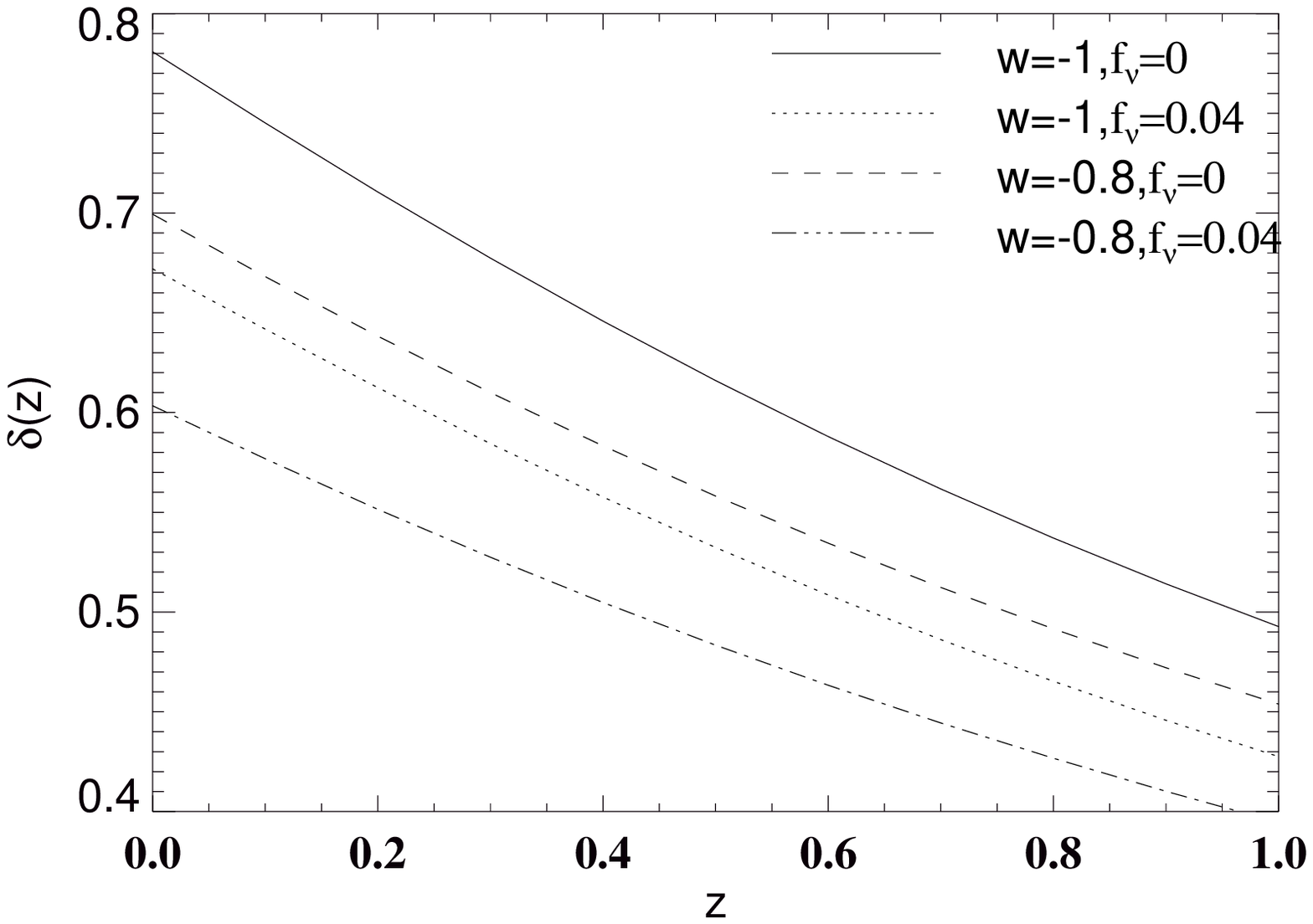}
\includegraphics[width=9cm,height=5cm]{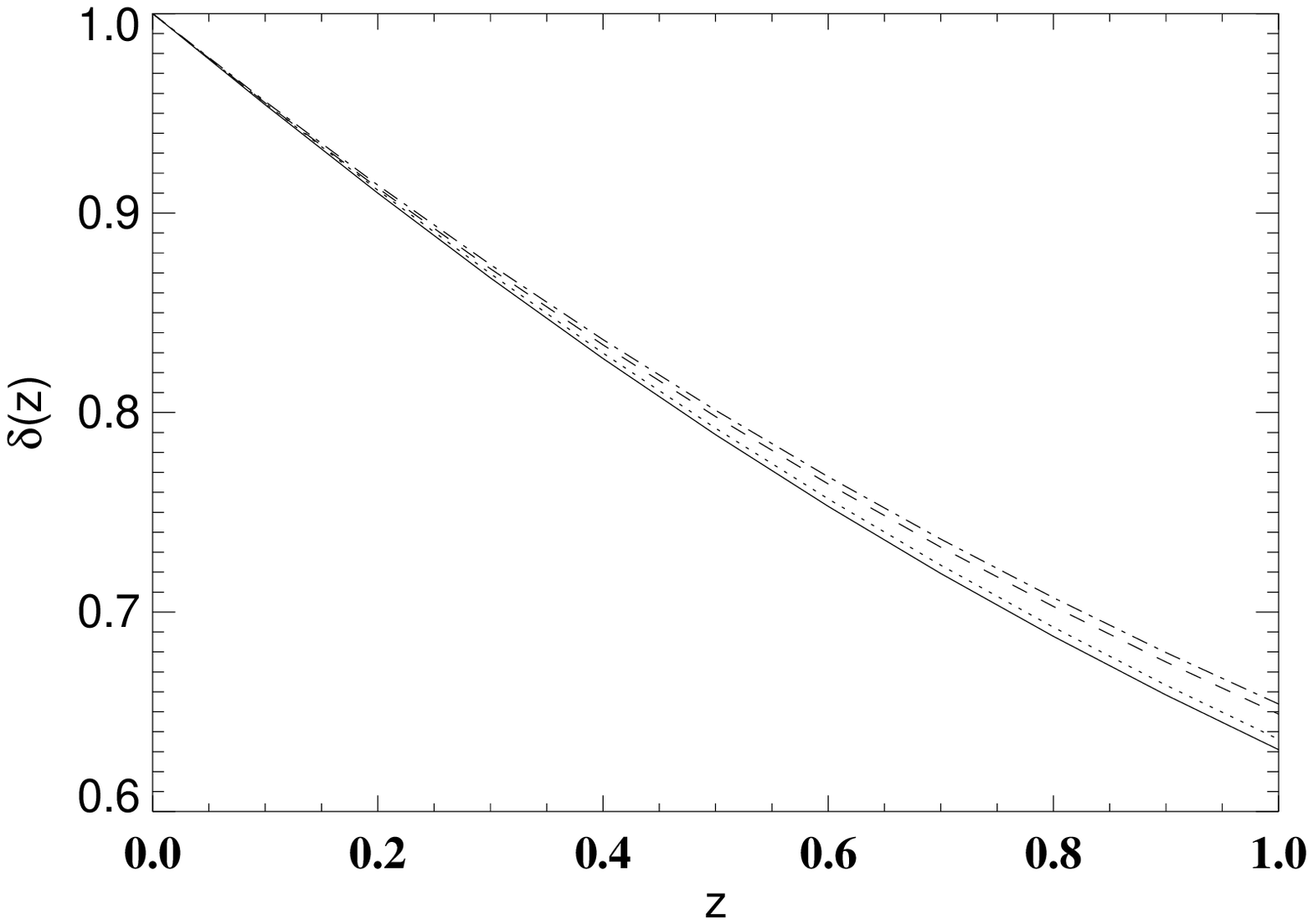}
\includegraphics[width=9cm,height=5cm]{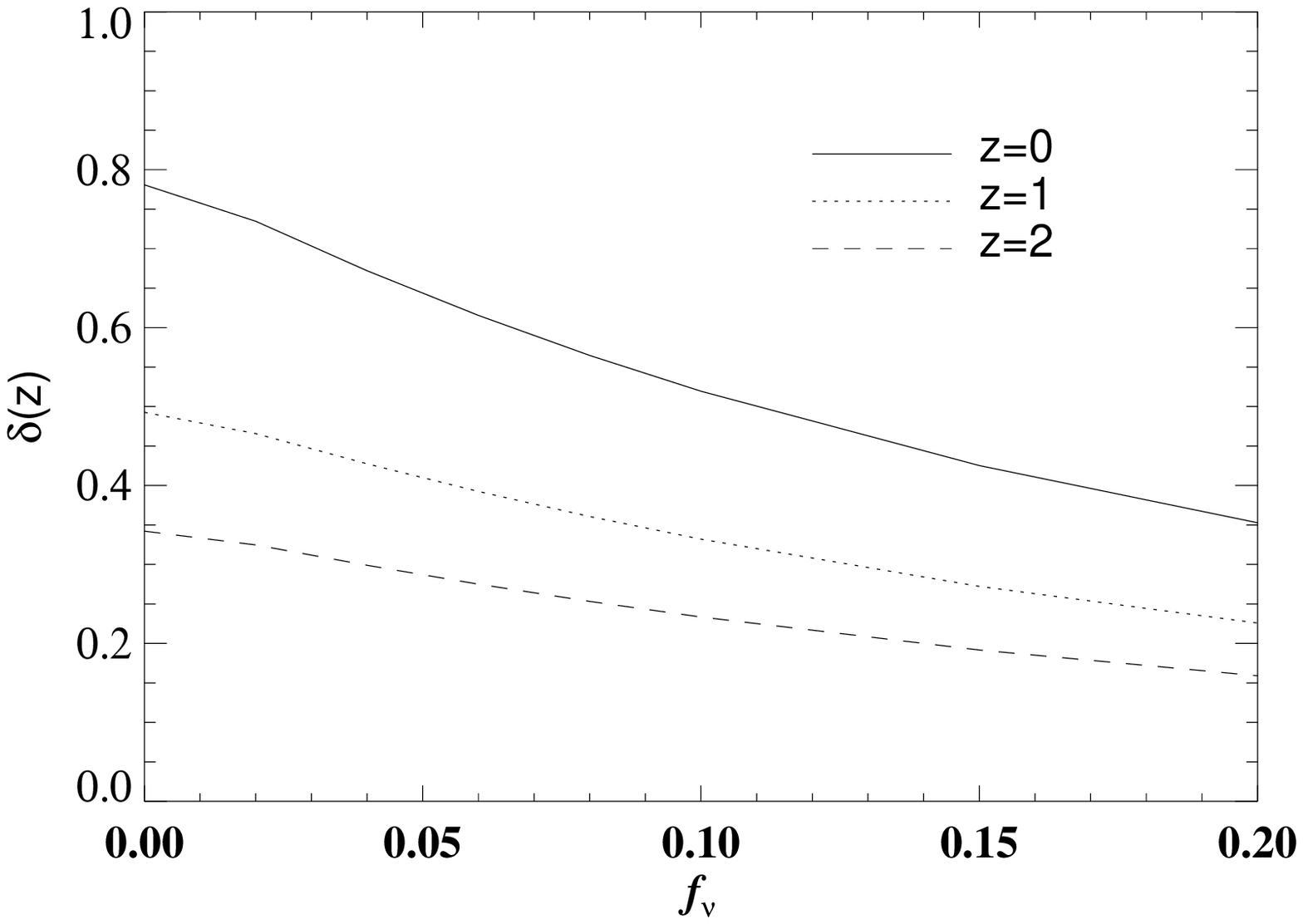}
\caption{\label{fig:fig4} Density fluctuation $\delta (z)$ on comoving 8$h^{-1}\textrm{Mpc} $ scale, normalized to CMB (top panel) and 
normalized to the value at $z=0$ (middle panel).  The models 
shown in the figure have $(w,f_\nu)$ equal to $(-1,0)$ (solid line), 
$(-1,0.04)$ (dotted line), $(-0.8,0)$ (dashed line), and 
$(-0.8,0.04)$ (dash-dotted line).  In the case of $w=-0.8$, we included dark energy perturbations according to CMBFAST.  We have assumed a spatially flat 
universe, adiabatic fluctuations, and fixed the matter density $\Omega_{m}=0.25$, 
baryon density $\Omega_{b}=0.04$, the Hubble constant $h=0.7$, scalar 
spectral index $n_{s}=1$, and the optical depth to reionization $\tau=0$. The lower plot shows the density fluctuation $\delta (z)$ as a function of $f_\nu$ at $z=0$ (solid line), $z=1$ (dotted line), and $z=2$(dashed line).
}
\end{figure}

\subsection{$\Lambda$ Dominated Universe}
\label{sec:lambda}
We now evaluate the linear growth of perturbation, equation (\ref{eq:eq2}), for a Universe with a cosmological constant.
On length scales below the present horizon dark energy does not cluster, and 
so it only affects $H=\dot{a}/a$.  For a flat cosmology with a dark energy 
component with constant equation of state $p=w\rho$, the Hubble parameter 
as a function of redshift $z$ is given by

\begin{equation}
H = \frac{\dot a}{a}=H_0 \sqrt{\Omega_{\rm m0}(1+z)^3 + (1-\Omega_{\rm m 0})(1+z)^{3(1+w)}},
\label{eq:hubble}
\end{equation}
where $H_0 = 100h\;{\rm km}\,{\rm s}^{-1}\,{\rm Mpc}^{-1}$ is the 
present value of the Hubble parameter, parametrized by the dimensionless 
Hubble parameter $h$, and $\Omega_{\rm m 0}$ is the present value of the 
matter density in units of the critical density that gives a 
spatially flat universe.  The linear growth factor $f$ is defined by \cite{matsubara}

\begin{equation}
f \equiv \frac{d\ln \delta}{d\ln a},
\label{eq:fgrow1}
\end{equation}
and equation (\ref{eq:eq2}) can be rewritten in terms of $f$ as 
\begin{equation}
\label{eq:second}
\frac{df}{d\ln a}=-f^2-\left[\frac{1}{2}-\frac{3}{2}w
(1-\Omega_{\rm m}(z))\right]f+\frac{3}{2}\Omega_{\rm m}(z)(1-f_\nu),
\end{equation}
where $a=(1+z)^{-1}$ is the scale factor of the universe, and $\Omega_{\rm m}(z) =H_{0}^{2} \Omega_{\rm m 0}(1+z)^3/H^{2}(z)$ is the time dependent density parameter of matter.  The Runge-Kutta integration of the set of equations (\ref{eq:fgrow1}), (\ref{eq:second}) simultaneously gives the growth factor and the logarithmic derivative of the growth factor. Equations (\ref{eq:fgrow1}), (\ref{eq:second}) are valid even when $w$ evolves with time. \par
Equation (\ref{eq:eq2}) is a simplified description of the effect 
of massive neutrinos on the growth of structures, but its solution is 
in good agreement with the results of detailed calculations with 
e.g. CAMB \cite{lewis1}.  We have checked this by comparing 
$\ln \delta$ found by solving (\ref{eq:second}) to 
the value obtained from CAMB (expressed there as 
$\sigma_8$, the 
root-mean-square mass fluctuation in spheres of radius $8\;h^{-1}\,{\rm Mpc}$). 
We note that 
the difference between the exact result and the solution of equation 
(\ref{eq:second}) is at most 7.5 \%.

Since the simple prescription employed above to describe the 
effect of neutrino masses on linear growth seems to work well over 
a range of parameters and cosmic epochs, we can motivate a 
simple analytical approximation to the linear growth factor $f$. 
In matter-dominated cosmologies this quantity is well approximated by 
$f\approx \Omega_{\rm m}^{0.6}(z)$ \cite{peebles,lightman}. 
An analytical approximation is also available for models 
with a cosmological constant, see e.g. \cite{lahav1}.  For models 
without massive neutrinos but with a more general dark energy component, 
Wang and Steinhardt \cite{wangstein} 
found that $f = \Omega_{\rm m}^{\alpha}(z)$, with 
\begin{equation}
\alpha=\alpha_{0}+\alpha_{1}\left[1-\Omega_{\rm m}(z)\right]
\label{eq:alpha}
\end{equation}
where
\begin{equation}
\alpha_0 = \frac{3}{5-\frac{w}{1-w}},
\label{eq:alpha_0}
\end{equation}
and
\begin{equation}
\alpha_1=\frac{3}{125}\frac{(1-w)(1-3w/2)}{(1-6w/5)^3},
\label{eq:alpha_1}
\end{equation}
for a constant $w$. We should note that $\alpha$ is a very weak function of 
redshift for $z > 1$. 

 To include the effect of massive neutrinos in the approximation for $f$, we note that for massive neutrinos the growth is not only redshift dependent but scale dependent as well. By using CAMB, we calculated $P(k,z)$ over a range of values for $f_{\nu}$, and varying $w$ and $\Omega_m$ in the proximity of $w=-1$ and $\Omega_m=0.25$, we find that a reasonable fit for a flat universe, $\Omega_m +\Omega_{DE}=1$, is given by  
 \begin{equation}
 f(z, k;f_\nu,w,\Omega_{\rm DE}) \approx \mu (k,f_{\nu}, \Omega_{DE})\Omega_{m}^{\alpha}(z),
 \label{eq:ffull}
 \end{equation}
 where
 \begin{equation}
 \mu (k,f_{\nu},\Omega_{DE})= 1-A(k)\Omega_{DE}f_{\nu}+B(k)f_{\nu}^{2}-C(k)f_{\nu}^{3}.
 \label{eq:mu}
 \end{equation}
 
\begin{table}
\begin{tabular}{|c|c|c|c|c|}
\hline
k (scales) & A(k) & B(k) & C(k) & $f$ \\
\hline
0.001 & 0 & 0 & 0 & 0.825265 \\
 0.01 & \,\,0.132 & 1.62 & 7.13 & 0.824272\\
 0.05 & 0.613 & 5.59 & 21.13 & 0.815508\\
 0.07 & 0.733 & 6.0 & 21.45 & 0.811596\\
  0.1   & 0.786 & 5.09 & 15.5 &  0.806680\\
  0.5  & 0.813 & \,\,0.803 & \,\,-0.844 & 0.789606\\
\hline

\end{tabular}
\label{table:fkall}
\caption{Numerical values for the coefficients A , B, C as a function of $k$ from equation (\ref{eq:mu}).$f$ for this values is given in the last column, for $f_{\nu}=0.08$, which corresponds to approximately $\Sigma m_{\nu}=1$eV. For $k>0.5 h\; \textrm{Mpc}^{-1}$, $\mu$ is better approximated using equation (\ref{eq:fsimple}).}
\end{table}

\begin{figure}[hbtp]
\includegraphics[width=9cm,height=5cm]{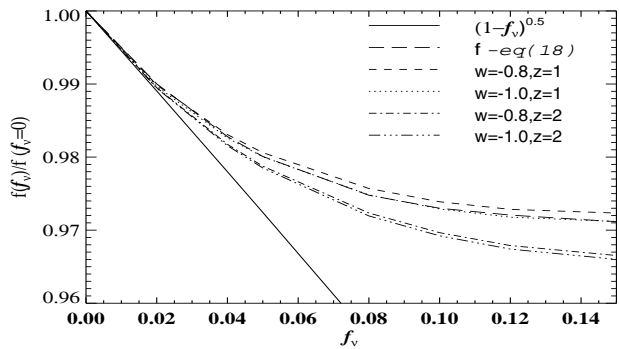}
\caption{Variation of the logarithmic growth factor $f=\frac{d\ln \delta}{d \ln a}$ with $f_{\nu}=\frac{\Omega_{\nu}}{\Omega_{m}}$ and with redshift at a fixed scale 8$h^{-1}\textrm{Mpc}$. The ratio $\frac{f(f_{\nu}> 0)}{f(f_{\nu}=0)}$ is shown as a function of $f_{\nu} $ for z=1 and z=2. he following examples are shown: $w=-0.8, z=1$(dashed line), $w=-1, z=1$(dotted line), $w=-0.8, z=2$(dash-dotted line) and $w=-1, z=2$(dash-triple dotted line).The solid line is $(1-f_{\nu})^{\alpha_0}$, from equation (\ref{eq:fsimple}), and the long-dashed line is the full polynomial found in equation (\ref{eq:fgrow5}), evaluated at $z=1$.Other parameters kept fixed $\Omega_{m}=0.3, \Omega_{b}=0.04, h=0.7$.}
\label{fig:fitfw}
\end{figure}

Numerical values for $A,B,C$ as a function of $k$ are given in Table II. They vary over the redshift range $z<10$ up to $1\%$.
Specifically for $8\;h^{-1}\; \textrm{Mpc}$ we recorded the $\sigma_8$ given by CMBFAST, and the resulting approximation is given as:
\begin{equation}
f(z;f_\nu,w,\Omega_{DE}) \approx (1-0.8\Omega_{DE}f_{\nu}+3.9f_{\nu}^{2}-9.8f_{\nu}^{3}) \Omega_{m}^{\alpha}(z).
\label{eq:fgrow5}
\end{equation}
This formula is valid for $f_{\nu}\leq 0.15$, and its accuracy is quite high for $w=-1, z=1$, see Figure \ref{fig:fitfw}. Performance at $w=-0.5$ or at $z=10$   is at most $2\%$ worse than the $w=-1, z=1$ case.

As expected $\mu =1$ on very large scales where $k < k_{nr}$.

On very small scales, where $k \gg k_{nr}$, we would expect that $f \approx [\Omega_m (1-f_{\nu})]^{\alpha_{0}}$ based on equation (\ref{eq:eq2}) and (\ref{eq:alpha_0}), ie.
\begin{equation}
 \mu = (1-f_{\nu})^{\alpha_0}. 
 \label{eq:fsimple}
 \end{equation}
 
 Another approximation on these small scales follows from equation (\ref{eq:julien}), and by \cite{takada}.
 However we note that this is based on the solution equation (\ref{eq:deltap}), which is only valid for an Einstein-de Sitter universe. We find that by solving the set of equations (\ref {eq:fgrow1}), (\ref{eq:second}) for a $\Lambda$ dominated universe, $\mu = 1-0.619f_{\nu}$, for $z=1, \Omega_m=0.25, w=-1$, compared to $\mu = p = 1-0.600 f_{\nu}$ of equations (\ref{eq:julien}),(\ref{eq:deltap}). We emphasize again this is only valid on very small scales, and for intermediate scales one should use equation (\ref{eq:mu}).

\begin{figure}[hbtp]
\includegraphics[width=9cm,height=6cm]{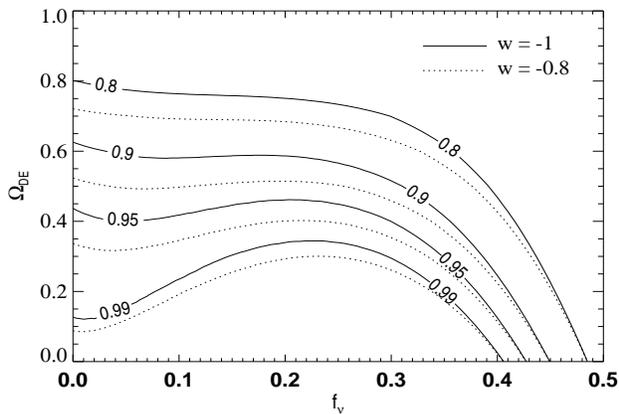}
\caption{Contour plot showing the degeneracy of $f_\nu$ and $\Omega_{DE}$ for $w=-1$ (full line) and $w=-0.8$ (dotted line), $h=0.7$, $\Omega_b=0.04$, at $z=1$ on 8 $h^{-1}\textrm{Mpc}$ scale. The contour levels correspond to values of $f$ using equation (\ref{eq:fgrow5}).}
\label{fig:fwcont}
\end{figure}

\section{The $w$-$f_\nu$ degeneracy}

Hannestad \cite{steendeg} pointed out that cosmological 
neutrino mass limits are considerably weakend if one allows for $w<-1$ 
(a case which is peculiar, as in this case the density will increase with the expansion of the universe). 
In his analysis, the parameter combination $\Omega_{\nu 0} h^2$ instead 
of $f_\nu$ was varied, and the explanation of the degeneracy he gave 
was as follows: since $f_\nu = \Omega_{\nu 0} / \Omega_{\rm m 0}$ 
determines the small-scale suppression of the matter power spectrum, 
one can compensate for a larger $\Omega_{\nu 0}$ by increasing 
$\Omega_{\rm m 0}$.  If one assumes a fixed $w=-1$ in the analysis, 
the Hubble diagram from supernovae Type Ia rule out values of $\Omega_{\rm m 0}$ much larger than $0.3$.  However, if one allows for $w<-1$, then the 
supernova data are compatible with considerably larger values of $\Omega_{\rm m 0}$, and hence a higher value of $\Omega_{\nu 0}$ can be accomodated by a 
given value of $f_\nu$. 
The degeneracy described above between $\Omega_{\nu 0}$ and $w$ is indirect.  We also emphasize again and again that the suppression is not just a function of 
$f_\nu$.

The degeneracy is most relevant when the {\it shape} 
of the matter power spectrum $P(k)$ at $z=0$ (or another fixed redshift) 
is used to constrain the neutrino mass (i.e. the bias between the 
galaxy distribution and total mass distribution is assumed constant and 
marginalized over).  Dark energy with a constant equation of state 
does not cluster on the scales probed by galaxy redshift surveys 
and does not affect the shape of the matter power spectrum, and hence 
$w$ affects the limit indirectly through the mechanism described above.     
Furthermore, the degeneracy is almost completely broken when using the  
WMAP 3-year data as opposed to the first year \cite{kristiansen}. \\

Based on the considerations in the previous subsections 
of this paper we would expect there to be a further and more direct 
degeneracy between $f_\nu$ and $w$: the same linear growth rate can 
result from a range of combinations of values of $w$ and $f_\nu$.  
In physical terms, this degeneracy can be understood as follows: 
neutrino free-streaming will supress growth of structure on small
scales.  
However, decreasing $w$ (for fixed 
$\Omega_{\rm m 0}$) prolongs the era of structure
formation,  
and reduces the value of the Hubble parameter in the matter-dominated
phase.  The Hubble parameter acts as a `friction term' in equation (\ref{eq:eq1}), so reducing it will enhance the linear growth factor, see equation (\ref{eq:hubble}). To sum up, one can at least partly compensate for increasing $f_\nu$ by decreasing $w$.    
This degeneracy would be relevant if one were to use the growth of 
structure to constrain $f_\nu$ and $w$.  In Figure (\ref{fig:fig4}) 
we illustrate this by plotting the root-mean-square mass fluctuation 
amplitude $\delta(z)$ as a function of redshift, 
for four different combinations of $w$ and $f_\nu$.  The middle
 panel 
shows the situation when if only the growth rate is measured (normalized at $z=0$): in 
that case distinguishing between the four cases will require 
very accurate measurements.  The situation is better
when the absolute values of the fluctuations are measured, as 
shown in the top panel of Figure (\ref{fig:fig4}).  This can 
be understood by noting that the absolute values depend, in 
the case when we normalize to the CMB on large scales, on $w$, 
but is only weakly dependent on $f_\nu$.  The reason for this is that
dark energy fluctuations 
are relevant on scales of the size of the horizon, and dominate  
integrated Sachs-Wolfe effect on the very largest scales. In contrast, the  
main effect of neutrinos on the CMB is on smaller scales through a small shift 
in the position of the peaks and a slight enhancement of their 
amplitude.  Thus, if the large-scale normalization is combined with 
a measurement of the growth rate, the degeneracy between $w$ and 
$f_\nu$ may  to a large extent be broken. Figure \ref{fig:fwcont} also shows the degeneracy of $f_{\nu}$ and $w$ through the linear growth factor $f$.

As indicated in Table \ref{table:geometry}, there is an interplay in the roles of neutrino mass 
and the DE equation of state, when  estimated observationally,  which depend
on geometry, the growth rate of perturbations and the shape of the power spectrum.
We discuss briefly two probes where results from this paper,
in particular the growth rate $f$  in equation (\ref{eq:fgrow5}), 
could help in understanding degeneracy.\\

The first is peculiar velocities (see e.g. \cite{dekel} for review).
The rms bulk flow is predicted in linear theory 
as: 
\begin{equation}
\langle v^2(R_*) \rangle =     
(2 \pi^2)^{-1}\; H_0^2 \;   
\int dk f^2 P(k) W_G^2(kR_*) 
\end{equation} 
where $W_G(kR_*)$ is a window function, 
e.g. $W(kR_*) = \exp(-k^2 R_*^2/2)$  for a Gaussian sphere
of radius $R_*$. 
The velocity field at low redshift is insensitive to geometry.
The power spectrum  $P(k)$ depends on the neutrino mass, but not on dark energy.
Massive neutrinos would suppress bulk flows \cite{elglahav}.
However in \cite{elglahav}  any dependence of $f$ on neutrino mass and dark energy was ignored.
Our equation (\ref{eq:fgrow5}) shows such dependence, which would result in extra suppression of the bulk flows and in some  degeneracy with $w$. This is also relevant on redshift distortion.

\begin{figure}
\includegraphics[width=7cm, height=9cm, angle=-90]{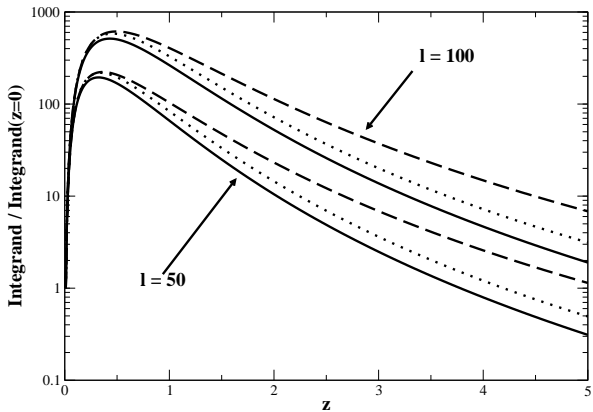}
\caption{The product $D^2(f-1)P$ in equation (\ref{eq:isw}), using equation (\ref{eq:fgrow5}) on 
$8h^{-1}\textrm{Mpc}$ scale, at $\ell=50$ (lower set of curves) 
and at $\ell=100$ (upper set of curves) for spatially flat universe models with $\Omega_{\rm m}=0.3$, $\Omega_{\rm b}=0.04$, $h=0.7$, $n_s=1.0$, and 
$w=-1$, $\Omega_\nu=0.0$ (full lines), $w=-1$, $\Omega_\nu = 0.01$ (dotted 
lines), $w=-0.8$, $\Omega_\nu = 0$ (dashed lines).
}
\label{fig:figISW}
\end{figure}

A second cosmological probe which depends on $f$ is the Integrated Sachs Wolfe effect
derived from cross-correlation of the CMB  with galaxy samples 
\cite{crittenden}.
In the small angle approximation 
the predicted spherical harmonic amplitudes are (e.g. \cite{afshordi,rassat}):
\begin{equation}
C_{gT}(\ell)=\frac{-3b_{\rm g}H_0^2\Omega_{m,0}}{c^3(\ell+1/2)^2}\int
{\rm d}  r\;\Theta(r)  D^2 H [f-1]P\left(\frac{\ell+1/2}{r}\right)
\label{eq:isw}
\end{equation}
where $\Theta(r)$ is a radial selection function for the galaxy survey, $b_g$ is the galaxy biasing factor,  and $D(t)$ is the  linear theory growth function.
The approximate relations for $f$, $D$ and $P(k)$ are useful to explore the degeneracies
between neutrino mass and dark energy.  As a simple illustration, we show 
in Figure (\ref{fig:figISW}) the product $D^2(f-1)P$ in the integrand above. 
We see that the selection function must extend to higher redshifts to 
distinguish between neutrino masses and $w\ne -1$. See \cite{julgast} for further analysis on the ISW effect in the presence of massive neutrinos.

\section{Conclusions}

We have demonstrated that the popular heuristic formula for the linear theory 
suppression of the matter fluctuations by free-streaming neutrinos, 
$\Delta P(k) / P(k) \approx -8f_\nu$, is valid only on very small 
scales ($k>0.8 h/ \textrm{Mpc}$). However, it is not of practical use as this is in the strongly non-linear regime of matter clustering. We also provide a modified code for the E$\&$H fitting formula. The linear 
growth factor in models with both massive neutrinos and a dark energy 
component with equation of state parameter $w=$constant, has been 
calculated numerically, and we have provided a fitting formula 
justified by simple arguments.  Furthermore, we have explained the 
indirect degeneracy between $w$ and $m_\nu$ found in \cite{steendeg}, 
and argued that there is a further, more direct degeneracy between 
$w$ and $f_\nu$ when the linear growth rate of density perturbations 
is considered.  This degeneracy can, however, be lifted by measurements 
of the absolute scale of the mass fluctuations, for example from the 
CMB.

\acknowledgments

AK acknowledges support by the University of London Perren Studentship.
The work of {\O}E is supported by the Research Council of Norway, 
project numbers 159637 and 162830. 
OL acknowledges support by PPARC Senior Research Fellowship. Special thanks to Filippe Abdalla and Ole Host, for fruitful discussions.
\par

%%%%%%%%%%%%%%%%%%%%%%%%%%%%%%%%%

%%%%%%%%%%%%%%%%%%%%%%%%%%%%%%%%%
\end{document}